\documentclass[aps,prd,twocolumn,psfig,floats]{revtex4}
\begin{document}
\newcommand{\be}{\begin{eqnarray}}
\newcommand{\ee}{\end{eqnarray}}
\newcommand{\nn}{\nonumber}
\newcommand{\noi}{\noindent\vspace{0.1cm}}
\newcommand\del{\partial}
\def\la{\lambda}
\def\ga{\gamma}
\def\om{\omega}
\def\al{\alpha}                            
\def\d{\partial}
\def\Tr{ {\rm Tr} }           
\def\Re{ {\rm Re} }
\def\un{\underline{n}}
\def\um{\underline{m}}
\def\uo{\underline{o}}
\def\up{\underline{r}}
\def\Box{{\bf \sqcap\!\!\!\!\sqcup}}

\vspace{-2.5cm}
\be
\hspace{14cm}{\rm SUNY\!-\!NTG\!-\!03/05} \nn
\ee
\vspace{-1cm}

\title{Impossibility of spontaneously breaking local
symmetries and the sign problem}

\vspace{1.5cm}

\author{K. Splittorff}
\affiliation{Department of Physics and Astronomy, SUNY, Stony Brook, NY\,11794, USA}
%\email{split@tonic.physics.sunysb.edu}
\preprint{SUNY-NTG-03/05}
\date{\today} 

\begin{abstract}
Elitzur's theorem stating the impossibility of spontaneous breaking 
of local symmetries in a gauge theory is reexamined. The existing 
proofs of this theorem rely on gauge invariance as well as positivity
of the weight in the Euclidean partition function.
We examine the validity of Elitzur's theorem in gauge theories for
which the Euclidean measure of the partition function is not positive
definite. We find that Elitzur's theorem does not follow from gauge
invariance alone. We formulate a general criterion under which 
spontaneous breaking of local symmetries in a gauge theory 
is excluded. Finally we illustrate the results in an exactly solvable 
two dimensional abelian gauge theory.
\end{abstract}

\maketitle

\section{Introduction}

A direct method to gain exact and non-perturbative
information about gauge theories is available if the weight,
$\exp[-S]$, in the Euclidean partition function is positive. 
Suppose one can find an inequality between two physical 
quantities which holds for any field configuration. In that case this
inequality also holds in the full theory provided that the 
weight in the Euclidean partition function is positive.  
In Quantum Chromo Dynamics (QCD) this method has been applied to the
propagators of quark bilinears and the resulting QCD/Weingarten 
inequalities \cite{QCDineq} have been successful 
in explaining aspects of the observed hadron mass spectrum. 
Another notable example 
of the use of positivity is the Vafa-Witten theorem \cite{VW}. However,
as already noted by Vafa and Witten in the original work \cite{VW}, the 
assumed positivity of the measure is not just a technical convenience
for the proof, it is actually a necessity: Vector-like symmetries can be
broken in gauge theories with a non-positive measure. For example the 
vectorial flavor symmetry in QCD is not protected by the Vafa-Witten 
theorem at non-zero $\theta$-angle. Theories with non-positive 
Euclidean weights, ie. theories with a sign problem, are not only of 
academic interest. For instance,
$\exp(-S)$ is complex in QCD at non-zero baryon chemical potential.

One of the central properties of gauge theories is that local
symmetries such as the local gauge symmetry itself can not break
spontaneously \cite{Elitzur};  

\noindent
{\bf Elitzur's Theorem:} 
{\sl In a gauge invariant theory a local
quantity with vanishing mean value on its orbit under the action of
the gauge group has zero ground state expectation value.}    

\noindent
The physical interpretation of the theorem is that spontaneous breaking
of the gauge invariance can only occur after having broken the
local symmetry explicitly. This is indeed what takes place in the
ordinary Higgs mechanism: First one chooses a gauge and in this gauge
the remaining global gauge symmetry is spontaneously broken.

In the proof of his theorem Elitzur implicitly assumed that the Euclidean
measure of the partition function was positive and explicitly made use of 
this. In general the 
proofs \cite{AFG,AFGM,K,FMS} of Elitzur's theorem are all based on the fact
that inequalities which hold for any field configuration continue to 
hold after integrating with respect to a positive measure.

Here we consider the validity of Elitzur's theorem in gauge theories
for which the Euclidean measure is not positive. 
We will investigate whether the assumed positivity of the measure is 
just a technical 
convenience or if it is essential for the theorem. That is, we
will examine whether gauge invariance alone is sufficient to protect
local order parameters from gaining a non-zero vacuum expectation 
value.    

The heart of the problem at hand is a double limiting process. First the
volume is taken to infinity and then the gauge variant source term is 
taken to zero. The discussion
is set in the framework of Euclidean lattice gauge theory. (For an elaborate
discussion of the relation between the lattice formulation of Elitzur's 
theorem and the continuum perturbative Higgs mechanism see \cite{FMS}.) 
To be 
specific, we follow Elitzur and consider the gauged planar spin model 
in $d$ dimensions. We expect, however, that the statements made 
generalize beyond this model.
In this context we establish that local gauge symmetry alone is not
sufficient to protect local symmetries from breaking spontaneously. 
We then formulate a new criterion under which the local symmetry
can not break: 
{\sl If all operators which are bounded, local, as well as gauge invariant
  have finite vacuum expectation values then Elitzur's theorem
  holds.} 
It is then shown that gauge theories with a positive weight
automatically satisfy this criterion. 

To illustrate how this criterion works we examine a two dimensional 
pure glue $U(1)$ theory which allows for analytic evaluation. Taking the
coupling to scale with the size of the system it is shown that the local
gauge invariance break spontaneously. However, in accordance with the 
criterion we show that the vacuum expectation value of a plaquette 
in this theory is infinite.

The organization of this paper follows the line of thought above. In
section \ref{sec:GPM} we define the lattice framework in which we will 
work. Then in section \ref{sec:example} we give a specific example
which illustrates why Elitzur's theorem does not follow from gauge
invariance alone. This example is then generalized in section 
\ref{sec:criterionI} and the general criterion under which Elitzur's
theorem holds is formulated. Finally, in section \ref{sec:U(1)pureglue} 
we show analytically how the criterion in a pure glue $U(1)$ theory in two  
dimensions excludes spontaneous breaking of local gauge invariance.

\section{The gauged planar spin model}
\label{sec:GPM}

In order to address the fate of local gauge invariance we  
will consider generalized versions of the gauged planar
spin model. The spin is parametrized through the angular field
$\phi$ 
\be
\left(\begin{array}{c} \cos(\phi_i)\\\sin(\phi_i)\end{array}\right) .
\ee
As indicated by the index $i$ the angular field is defined on the sites
of a space-time lattice. The gauge field $A_{i,\um}$ lives on the
link from the site $i$ in the direction $\um$ and is also an angular
variable. The local gauge transformation of the fields is  
\be\label{localgauge}
\phi_i & \to &\phi_i + C_i \\ 
A_{i,\um} & \to & A_{i,\um} + C_i - C_{i+\um} \nn  .
\ee
Here $C_i$ is a function taking arbitrary complex values.

The action considered by Elitzur \cite{Elitzur} is 
\be
\label{action-E}
& S & =K\sum_{i=1}^N\sum_{\um=1}^d\cos(\phi_i-\phi_{i+\um}-A_{i,\um})  \\
  & + & \frac{1}{g^2}\sum_{i=1}^N\sum_{\un,\um=1;\un\neq
  \um}^d \hspace{-4mm} 
\cos(A_{i,\un}+A_{i+\un,\um}-A_{i+\um,\un}-A_{i,\um})  ,\nn
\ee
where $N$ is the volume, $d$ is the number of dimensions, and
$K$ and $g$ are constants (for further discussion of this model 
see \cite{kogut}). This action is by
construction invariant under the local gauge transformation
(\ref{localgauge}). Besides being gauge invariant the action is 
real and periodic in the fields. Because the action is real the weight
$\exp[-S]$ in the partition function   
\be
Z(N,J=0) & = & \int_{-\pi}^\pi \prod_{i=1}^N{\rm d}
  \phi_i\prod_{\um=1}^d {\rm d}A_{i,\um} e^{-S(\phi,A)}   
\ee
is positive. In this letter we will consider general gauge invariant
actions of the angular fields $\phi_i$ and $A_{i,\um}$.

The evaluation of a vacuum expectation value involves a double limit. 
First one introduces a source $J$ for an external field $F(\phi,A)$. 
The vacuum expectation value of an operator $O(\phi,A)$ is then 
defined as 
\be \label{genvev}
&& \langle O(\phi,A) \rangle \equiv \lim_{J\to0}\lim_{N\to\infty}\\
&& \frac{\int_{-\pi}^\pi
\prod_{i=1}^N{\rm d} \phi_i \prod_{\um=1}^d {\rm d}
A_{i,\um}  e^{-S(\phi,A)} O(\phi,A) e^{J F(\phi,A)}}{Z(N,J)} .\nn
\ee
We are interested in the vacuum expectation value of a gauge variant 
local operator whose average over the gauge orbit is zero. 
{\sl Local} means that $O(\phi,A)$ only depends on fields at a finite
number of sites and links. As a  
further restriction we shall only consider a source which vanishes 
upon averaging over the gauge orbit.   
The external source will be chosen to break the local gauge invariance
explicitly. Having an external magnetic field in mind we will for example 
choose 
\be\label{source}
F(\phi,A)=\sum_{i=1}^{N}\sum_{\um=1}^{d}\cos(A_{i,\um})  .
\ee
With the action (\ref{action-E}) and this source 
Elitzur showed that for a fixed link $(j,\un)$
the vacuum expectation value of $\cos(A_{j,\un})$ defined by 
\be
&& \langle \cos(A_{j,\un}) \rangle 
 \equiv  \lim_{J\to0}\lim_{N\to\infty} \\
&& \frac{\int_{-\pi}^\pi
\prod_{i=1}^N{\rm d} \phi_i \prod_{\um=1}^d {\rm d}
A_{i,\um}  e^{-S(\phi,A)} \cos(A_{j,\un}) e^{J F(\phi,A)}}{Z(N,J)} \nn
\ee
vanishes. The choice of $O(\phi,A)\equiv \cos(A_{j,\un})$ is not
essential. However, it must be a function which vanishes upon average over
its gauge orbit.  

The original paper by Elitzur is very clear and rather than
repeating the proof we encourage the reader to consult Elitzur's
original paper \cite{Elitzur}. Below we will give an alternative and
general proof of Elitzur's theorem when the weight is positive.     

{\sl Assumptions:} Here we will consider general actions 
describing the angular fields $\phi_i$ and $A_{i,\un}$. The action is 
assumed to be gauge invariant but the weight $\exp[-S(\phi,A)]$ is  
not necessarily real and positive. We will only consider functions 
of the angular fields $\phi_i$ and $A_{i,\un}$ which are periodic on 
$[-\pi,\pi]$. Moreover, we will only consider actions and sources that 
act locally, i.e. where the individual terms only connect nearby sites 
and links. 
Within these assumptions our main goal is to find a set of 
constraints under which Elitzur's theorem holds.

\section{Sensitivity to local probes}
\label{sec:example}

In this section we consider a specific choice of the 
external field and the gauge variant local operator. 
We state the conditions under which the limits $N\to\infty$ 
and $J\to0$ in (\ref{genvev}) do not commute. The example 
is generalized in the following section.
\noi

We evaluate the vacuum expectation value (vev) of ($j$ is a fixed site 
and $\un$ is a fixed direction)
\be
\cos(\phi_j-\phi_{j+\un})
\ee
with the source given in (\ref{source}). That is
\cite{footnote}
\be
&& \langle \cos(\phi_j-\phi_{j+\un}) \rangle \equiv 
\lim_{J\to0}\lim_{N\to\infty} \\
&&\frac{\int_{-\pi}^\pi
{\rm d} \{\phi\} {\rm d}\{A\} e^{-S(\phi,A)} 
e^{J \sum_{i,\um}\cos(A_{i,\um})} \cos(\phi_j-\phi_{j+\un})}{Z(N,J)} . \nn
\ee
In order make this evaluation we first change variables from 
$(\phi_i,A_{i,\um})$ to $(\phi_i,l_{i,\um})$ where
\be
l_{i,\um} \equiv \phi_i-\phi_{i+\um}-A_{i,\um}  .
\ee
The action is only a function of $l_{i,\um}$ due to gauge invariance. 
Furthermore, the Jacobian is field independent and cancels between
numerator and denominator when evaluating the vev
\begin{widetext}
\be
\langle \cos(\phi_j-\phi_{j+\un}) \rangle & \equiv &  
\lim_{J\to0}\lim_{N\to\infty} \frac{\int  {\rm d} 
\{l\} e^{-S(l)} \int 
{\rm d}\{\phi\}   e^{J
  \sum_{i,\um}\hspace{-2mm}\cos(l_{i,\um}-\phi_i+\phi_{i+\um})}
\cos(\phi_j-\phi_{j+\un})}{\int_{-\pi}^\pi {\rm d} \{l\} e^{-S(l)}
  \int_{-\pi}^\pi    
{\rm d}\{\phi\}  e^{J \sum_{i,\um}\cos(l_{i,\um}-\phi_i+\phi_{i+\um})}}  . 
\ee
Because of the choice of the source the denominator does not have a 
term linear in $J$.
Expanding $\exp[J\sum_{i,\um} \cos(l_{i,\um}-\phi_i+\phi_{i+\um})]$ in
the numerator we have  
\be
&& \int_{-\pi}^\pi  {\rm d}\phi_j {\rm d}\phi_{j+\un}
e^{J\sum_{i,\um}\cos(l_{i,\um}-\phi_i+\phi_{i+\um})}
\cos(\phi_j-\phi_{j+\un}) =  J 2\pi^2 \cos(l_{j,\un}) + {\cal O}(J^2).   
\ee
Hence for small $J$ we find
\be\label{finalgeneral}
&& \langle \cos(\phi_j-\phi_{j+\un}) \rangle  =  
\lim_{J\to0}\lim_{N\to\infty}
\frac{J/2 \int_{-\pi}^\pi \prod_{i,\um} {\rm d}l_{i,\um} e^{-S(l)}   
\cos(l_{j,\un})}{\int_{-\pi}^\pi \prod_{i,\um}
{\rm d} l_{i,\um} e^{-S(l)}} + {\cal O}(J^2).   
\ee
This shows that coherence (ie. gauge invariant modification) in the
numerator is possible even if it does not happen in the denominator.
\end {widetext} 

{\sl If} $e^{-S(l)}$ is a {\sl real and positive} function then we can
use that $|\cos(l_{j,\um})|\leq 1$ to show that for all $N$ and $J$ 
\be
\frac{J/2 \int_{-\pi}^\pi \prod_{i,\um} {\rm d}l_{i,\um} e^{-S(l)}   
\cos(l_{j,\un})}{\int_{-\pi}^\pi \prod_{i,\um}
{\rm d} l_{i,\um} e^{-S(l)}} & \leq & \frac{J/2 Z(N,J)}{Z(N,J)}. \nn
\ee
The $N\to\infty$ limit on the right hand side is trivial and the leading term
in $J$ of $\langle \cos(\phi_j-\phi_{j+\un}) \rangle$ vanishes linearly with
$J$.  The same is true for higher order terms in $J$,
thus confirming Elitzur's theorem provided that $e^{-S(l)}$ is
positive.  
\noi 

{\sl If} $e^{-S(l)}$ is {\sl not positive} we can not draw such a
conclusion.    
Let us go back to (\ref{finalgeneral}) and first perform the integral
over all $l_{i,\um}$ with $i\neq j$ and $\um\neq\un$. This leaves
\be\label{vevalmostint}
\langle \cos(\phi_j-\phi_{j+\un}) \rangle = 
\lim_{J\to0}\lim_{N\to\infty} \hspace{3cm} \\ \ \ \ \ \ 
\frac{J/2 \int_{-\pi}^\pi {\rm
d} l_{j,\un} 
\cos(l_{j,\un})f(l_{j,\un},N)}{\int_{-\pi}^\pi {\rm d} l_{j,\un} 
f(l_{j,\un},N)} + {\cal O}(J^2) ,   \nn
\ee
where
\be
f(l_{j,\un},N) \equiv \int_{-\pi}^\pi \prod_{i\neq j, \um\neq\un}{\rm
d} l_{i,\um} e^{-S(l)} . 
\ee
The only requirement we have imposed on the function $f(l_{j,\um},N)$
is that it is a periodic function in $l_{j,\um}$. Hence gauge
invariance and periodicity does not exclude that, say, $
f(l_{j,\un},N) = 1/N+\cos(l_{j,\un})$.
In this case the $N\to\infty$ limit in (\ref{vevalmostint}) is infinite; 
the numerator being larger by a factor of $N$ than the
denominator. Therefore, the limits $N\to\infty$ and $J\to 0$ do 
not commute, signaling a possible non-trivial vev of 
$\cos(\phi_j-\phi_{j+\un})$.   
This example illustrates that gauge invariance alone is not sufficient
to prevent local order parameters from obtaining a non-zero vev. The
object is now to formulate a constraint on $e^{-S(l)}$ which is less
restrictive than positivity but nevertheless allows us to exclude
spontaneous breaking of local symmetries in a gauge theory.
\vspace{-0.4cm}

\section{Criterion Under Which Elitzur's Theorem Holds}
\label{sec:criterionI}
\vspace{-0.1cm}

We now generalize the example in the previous section and give a 
general criterion for when Elitzur's theorem holds. 
Consider a local, bounded, periodic, and gauge variant function, 
$O(\phi,A)$, which vanishes on average over its gauge orbit and consider 
a general bounded and periodic source $F(\phi,A)$.
Starting from the definition (\ref{genvev}) and expanding in $J$ we get
\be\label{finalevenmoregeneral}
\langle O \rangle = \lim_{J\to0}\lim_{N\to\infty} \hspace{5cm} \\
\frac{J \int_{-\pi}^\pi \prod_{i,\um} {\rm
    d}l_{i,\um} e^{-S(l)}    
g(l)}{\int_{-\pi}^\pi 
{\rm d} \{l\} e^{-S(l)}+J\int_{-\pi}^\pi
{\rm d} \{l\} e^{-S(l)}F(l)} + {\cal O}(J^2)\nn   
\ee
where 
\be
g(l) & \equiv & \int_{-\pi}^\pi \prod_{i=1}^N {\rm d}\phi_i F(\phi,l)
O(\phi,l),   \\ 
F(l) & \equiv & \int_{-\pi}^\pi \prod_{i=1}^N {\rm d}\phi_i F(\phi,l). \nn
\ee
In the numerator the term of order $J^0$ vanish identically when integrating
over $\phi$ since $O$ vanish on average over its gauge orbit.
For the same reason and because $F$ acts locally the function $g(l)$ must be 
gauge invariant, periodic, bounded, and can only depend on $l_{i,\um}$ 
belonging to a finite part of the lattice. 

Now, provided that 
\be
\label{Fvev0}
\lim_{N\to\infty} \frac{\int_{-\pi}^\pi\prod_{i,\um}{\rm d} l_{i,\um}
  e^{-S(l)}}{\int_{-\pi}^\pi \prod_{i,\um} {\rm d} l_{i,\um} e^{-S(l)}F(l)}
=0,   
\ee
we can drop the first term in the denominator of
(\ref{finalevenmoregeneral}). In that case the factors of $J$ will cancel 
and the expectation value is given by
\be
\langle O \rangle = \lim_{N\to\infty} 
\frac{\int_{-\pi}^\pi \prod_{i,\um} {\rm d}l_{i,\um} e^{-S(l)}    
g(l)}{\int_{-\pi}^\pi
\prod_{i,\um} {\rm d}l_{i,\um} e^{-S(l)}F(l)}.
\ee
With the factor of $J$ canceling explicitly the vev is potentially non-zero. 
In section \ref{sec:U(1)pureglue} we construct an example 
where $g$ and $F$ are identical and the vev is thus unity. In the example 
where $F(l)=0$ given in the previous section we required that 
\be
\label{gvev0}
\lim_{N\to\infty} \frac{\int_{-\pi}^\pi\prod_{i,\um}{\rm d} l_{i,\um}
  e^{-S(l)}}{\int_{-\pi}^\pi \prod_{i,\um} {\rm d} l_{i,\um} e^{-S(l)}g(l)}
=0
\ee
in order to get a non-zero vev.
The property (\ref{Fvev0}) or (\ref{gvev0}), which is needed in order to 
obtain a non-zero value of $\langle O\rangle$, has a direct physical 
meaning: The ratios are by definition the inverse vev's of
$F$ and $g$ respectively measured without any external source.
Now, since $g$ and $F$ are arbitrary, periodic, bounded, and gauge invariant 
functions we can formulate \\ 
{\bf The general criterion:} {\sl
If all bounded, local, and gauge invariant operators have finite vacuum 
expectation values when measured without an external source 
then spontaneous breaking of local symmetries is excluded.} 

We expect that this criterion is valid beyond the present abelian planar spin
models considered here. Proving this, however, is not trivial as soon as the
integrations become non-compact.

\noi

An alternative way to formulate the criterion is by considering the 
Fourier expansion of the weight
\be\label{FourierExp-S}
&& \exp[-S(l)]= \\
&& a_0 + \sum_{i,\um }a^{(1)}_{i,\um}\cos(l_{i,\um}) +
\sum_{i,\um}a^{(2)}_{i,\um}\cos(2l_{i,\um})  +
\ldots \nn \\
&& + \sum_{i,\um} b^{(1)}_{i,\um}\sin(l_{i,\um}) +
\sum_{i,\um}b^{(2)}_{i,\um}\sin(2l_{i,\um}) +\ldots .\nn
\ee
In the partition function the integration over the field
dependent terms vanishes. That is 
\be
a_0(N)=Z(N,J=0)/(2\pi)^N  .
\ee
The other Fourier coefficients are the vacuum expectation values of 
the Fourier modes at zero external source.
Therefore, in terms of the  Fourier expansion, the criterion for establishing
Elitzur's theorem is: 

\noindent
{\sl If the ratio of all Fourier coefficients of $\exp[-S]$ and the 
constant mode is finite for $N\to\infty$ then the vev of a local 
quantity which vanishes in average over its gauge orbit is zero.} 
\noi

In order for the Fourier expansion of $\exp[-S]$ to be convergent the
coefficients 
must be finite in the $N\to\infty$ limit. Hence, a necessary
requirement for breaking Elitzur's theorem is that $Z(N,J=0)$ is
zero in the $N\to\infty$ limit. This, however, is quite natural; 
$Z(N,J=0)$ will normally be the generating functional for some
extensive quantity. Consider for example the baryon density in QCD;
there we will expect that $Z(N,J=0)\propto \exp[-\mu_B^2 N]$ where 
$\mu_B$ is the baryon chemical potential (see eg. \cite{HJV}). 

\noi 

In the next section we study a two dimensional U(1) model and show 
analytically 
how the criterion excludes spontaneous breaking of local symmetries. 
First, however, let us show that the criterion is fulfilled
automatically if $\exp[-S]$ is positive. 

\noi

{\sl Positivity revisited:} Assuming that $\exp[-S(l)]$ is positive 
we have for any bounded function with max$|f(l)|=f_{max}>0$
\be
\frac{\int_{-\pi}^\pi\prod_{i,\um} {\rm d} l_{i,\um}
e^{-S(l)}}{\int_{-\pi}^\pi\prod_{i,\um} {\rm d} l_{i,\um}
e^{-S(l)}f(l)} \geq \frac{1}{f_{max}}>0.
\ee
Hence by the above criterion spontaneous breaking of local symmetries 
is excluded.

As for the Fourier coefficients it is trivial to show
that the constant mode $a_0$ is larger than all other modes if 
$\exp[-S(l)]$ is positive. 
That is, the amplitudes of the oscillatory terms as compared to 
the constant term are restricted by the positivity of the
measure. In particular the possibility that the ratio can 
be infinite for $N\to\infty$ is excluded and thus $\langle
O(\phi,A)\rangle = 0$. This reestablishes Elitzur's theorem for a
positive weight in a general framework.

\section{$U(1)$ gauge theory in 2 dimensions}
\label{sec:U(1)pureglue}

In order to make the general discussion from the previous sections more 
concrete we now look at a pure glue $U(1)$ gauge theory in 2 dimensions. 
This theory is 
analytically solvable even when the action is supplemented
by an imaginary term. To be specific we will consider the weight
\be
e^{-S(A)} =
e^{-\beta\sum_{i,\um,\un}
  \cos(\Box_{i,\um,\un})}e^{-i 2 \sum_{i,\um\in L} A_{i,\um}} ,    
\ee
where the plaquette is defined as
\be
\Box_{i,\un,\um}\equiv A_{i,\un}+A_{i+\un,\um}-A_{i+\um,\un}-A_{i,\um}  
\ee
and $L$ defines the contour of a Wilson loop. 
This complex weight has been used previously \cite{Langevin} to discuss 
the Langevin formulation of Monte Carlo simulations on a complex weight.

We chose to measure the vacuum expectation value of 
\be
O(A)\equiv\exp[i2(A_{j,\uo}+A_{j+\uo,\up})] 
\ee
in the presence of the source 
\be
F(A) \equiv  \exp[i2\Box_{j,\uo,\up}]
           + \exp[i2(-A_{i+\up,\uo}-A_{i,\up})].
\ee
The site $j$ and the directions $\uo$ and $\up$ are fixed so that the
plaquette $\Box_{j,\uo,\up}$ lies inside the contour $L$ of the Wilson loop.
Note that, the orientation of this plaquette is chosen opposite of that of
the Wilson loop in the weight. 
We will consider extremely strong coupling, $\beta\sim\frac{1}{N}$, and
show that the local invariance, $A_{i,\un}\to A_{i,\un} + C_i- C_{i,\un}$, is 
spontaneously broken. Then we show that in this limit $\langle \exp[i2\Box]
\rangle =\infty$ in accordance with the general criterion formulated above. 
We proceed as we did in section \ref{sec:example} by choosing gauge invariant
coordinates (now the plaquettes) and integrating over the remaining variables.
This is possible because there are twice as many links as there are
plaquettes in two dimensions. To be specific we change coordinates according
to 
\be
\left(\begin{array}{c}A_{i,\un} \\ A_{i+\un,\um} \\ A_{i+\um,\un} \\
    A_{i,\um}\end{array}\right) \to
  \left(\begin{array}{c}A_{i,\un}+A_{i+\un,\um}+A_{i+\um,\un}+A_{i,\um} \\
                        A_{i,\un}+A_{i+\un,\um}+A_{i+\um,\un}-A_{i,\um} \\ 
                        A_{i,\un}+A_{i+\un,\um}-A_{i+\um,\un}-A_{i,\um}  \\
                        A_{i,\un}-A_{i+\un,\um}-A_{i+\um,\un}-A_{i,\um} 
\end{array}\right). \nn
\ee
Note that, the third coordinate simply is the plaquette. The other three
coordinates are not gauge invariant. Due to the periodicity of the integrand 
the integration range on each of the new coordinates remains the interval 
$[-\pi,\pi]$. Expanding in the source $J$ yields 
\begin{widetext} 
\be
 \langle e^{i2(A_{j,\uo}+A_{j+\uo,\up})}\rangle 
& \equiv & \lim_{J\to0}\lim_{N\to \infty} 
\frac{\int {\rm d}\{A\} e^{-S(\Box)+J\exp[i2\Box_{j,\uo,\up}]}
  e^{J\exp[i2(-A_{i+\up,\uo}-A_{i,\up})]}e^{i2(A_{j,\uo}+A_{j+\uo,\up})}}
{\int {\rm d}\{A\} e^{-S(\Box)+J\exp[i2\Box_{j,\uo,\up}]}
  e^{J \exp[i2(-A_{i+\up,\uo}-A_{i,\up})]}} \nn  \\
& = & \lim_{J\to0}\lim_{N\to \infty} 
\frac{J\int {\rm d}\{\Box\} e^{-S(\Box)}e^{i2\Box_{j,\uo,\up}}}
{\int {\rm d}\{\Box\} e^{-S(\Box)}+J\int {\rm d}\{\Box\}
  e^{-S(\Box)}e^{i2\Box_{j,\uo,\up}}}+{\cal O}(J^2).
\label{U(1)}
\ee
If the second term in the denominator dominates the first then 
$\langle \exp[i2(A_{j,\uo}+A_{j+\uo,\up})]\rangle = 1$. That is, 
the vev of a local and gauge variant operator which vanishes on average 
over its gauge orbit is non-zero.
Since the gauge group is abelian and we consider two dimensions
the two terms in the denominator can be evaluated analytically \cite{BDI}.
Using that for abelian theories we have
\be
e^{i\sum_{i,\um\in L} A_{i,\um}}= e^{i\sum_{\Box_{i,\um,\un}\in L}
  \Box_{i,\um,\un}}    
\ee
one gets ($A$ is the total area and $A_L$ is the area of the Wilson
loop $L$) 
\be 
\int {\rm d}\{\Box\} e^{-S(\Box)} & = & 
 \left(\int_{-\pi}^\pi {\rm d}\Box  
e^{-\beta \cos[\Box]}e^{-i2\Box}\right)^{A_L}
\left(\int_{-\pi}^\pi {\rm d}\Box  
e^{-\beta \cos[\Box]}\right)^{A-A_L} \\\
\int {\rm d}\{\Box\} e^{-S(\Box)}e^{i2\Box_{j,\uo,\up}} & = & 
\left(\int_{-\pi}^\pi 
{\rm d}\Box  
e^{-\beta \cos(\Box)}e^{-i2\Box}\right)^{A_L-1}
\left(\int_{-\pi}^\pi {\rm d}\Box  
e^{-\beta \cos(\Box)}\right)^{A-A_L+1}. \nn
\ee
\end{widetext}
Hence, the ratio is simply a ratio of modified Bessel functions
\cite{BDI,Langevin} 
\be
\frac{ \int \! {\rm d}\{\Box\}
  e^{-S(\Box)}e^{i2\Box_{j,\uo,\up}}}{\int {\rm d}\{\Box\}
  e^{-S(\Box)}}
 \!= \!\frac{\int {\rm d}\Box e^{-\beta \cos(\Box)}}
{\int \!  {\rm d}\Box e^{-\beta \cos(\Box)}e^{-i2\Box}}
 \!= \! \frac{I_0(\beta)}{I_2(\beta)}. \label{vevexpi2Box} 
\ee
For small values of $\beta$ this ratio diverges like $8/\beta^2$. 
Consequently, we can
neglect the first term in the denominator of (\ref{U(1)}) provided that
$\beta^2/J \ll 1$ in the limits $N\to\infty$ followed by $J\to0$. For 
example, with $\beta = 1/N$ we find that 
$\langle\exp[i2(A_{j,\uo}+A_{j+\uo,\up})]\rangle = 1$. We emphasize that 
this is possible due to the complex nature of the weight and not just because
we allow $\beta$ to be of order $1/N$. 
(If $S(\Box)$ is real then the first term in the denominator of (\ref{U(1)})
is larger than the second term for all $J<1$.) In the evaluation above we 
have only kept track of the leading terms in $J$. This was done in order to
keep the form of the equations as close to those of the previous sections. 
We stress, however, that it is possible to evaluate
$\langle\exp[i2(A_{j,\uo}+A_{j+\uo,\up}]\rangle$, as given by  
the ratio in the first line if (\ref{U(1)}), for all 
values of $J$ and $\beta$. The result is 
\be
 \langle e^{i2(A_{j,\uo}+A_{j+\uo,\up})}\rangle 
& \equiv & \lim_{J\to0}\lim_{N\to \infty} 
\frac{J\sum_{k=0}^\infty (J^k/k!) I_{2k}(-\beta)}{\sum_{k=0}^\infty (J^k/k!)
  I_{2(k-1)}(-\beta)}. \nn\\  
\ee
With $\beta\propto 1/N$ in the limit $N\to\infty$ the $k=0$ term dominates 
in the numerator while the $k=1$ term dominates in the denominator. Therefore
\be
 \langle e^{i2(A_{j,\uo}+A_{j+\uo,\up})}\rangle 
& \equiv & \lim_{J\to0} 
\frac{J I_{0}(0)}{(J^1/1!) I_{0}(0)} =1
\ee
in agreement with what we found above.

In order to make the connection to the criterion formulated in the previous
section we finally consider the expectation value of 
$\exp[i2\Box_{j,\uo,\up}]$ on the same weight but with zero external 
source. This expectation value was evaluated in (\ref{vevexpi2Box}) where we
found that $\langle\exp[i2\Box_{j,\uo,\up}]\rangle$ diverges
like $1/\beta^2$. From this we conclude: in the case were 
$\langle \exp[i2(A_{j,\uo}+A_{j+\uo,\up})]\rangle=1$ we also have that 
the vev of a local, bounded, and gauge invariant  operator (measured
without an external source) is infinite. 
This example illustrates how the general criterion excludes local 
order parameters from getting a non-zero vev.

\section{Summary}

Elitzur showed that positivity of the measure and gauge invariance is
sufficient to protect local symmetries from breaking
spontaneously. Here we have considered abelian gauge theories with 
non-positive measures and have
found that gauge invariance alone is not sufficient to prevent
a spontaneous breaking of local symmetries. With a non-positive measure
the partition function can be dominated by delicate cancellations. 
We have formulated a general criterion under which Elitzur's theorem 
remains valid. It was then shown how in this formulation positivity of 
the measure implies the vanishing of local order parameters.
The restriction in the criterion on the weight $\exp[-S]$ was formulated 
in terms of the vacuum expectation values of local gauge invariant 
operators: 
{\sl If all bounded, local, and gauge invariant operators have finite vacuum
expectation values then Elitzur's theorem holds.}
Finally, we illustrated analytically how the criterion works in the 
case of a $U(1)$ pure glue theory in two dimensions.

Whether the restriction in the criterion is fulfilled for QCD at non-zero 
baryon chemical potential or other physically relevant field theories 
with a non-positive Euclidean weight is at present not clear. 
However, any theory which has infinite expectation values for 
bounded, local, and gauge invariant operators is likely to be ill-defined. 
For instance, the infinities of the local variables can imply that also 
thermodynamic quantities like the baryon density are infinite.    

We round off with two remarks:
In the generalized gauged planar spin theories considered above the 
$U(1)$ invariance allowed us to choose variables such that gauge
invariance was manifest. Such a change of variables is in general not
trivial to make. However, we expect that also in non-abelian
gauge theories one can set up a criterion under which Elitzur's
theorem holds. 

Finally, let us mention that L\"uscher \cite{Luscher} has constructed a
proof of Elitzur's theorem in the Hamiltonian formulation of lattice 
$U(1)$ and $SU(2)$ pure Yang-Mills theories. Perhaps that line of work
can be extended to a gauge theory with dynamical fermions and maybe
even to QCD at non-zero baryon chemical potential. Such an extension
may cast light on additional physical constraints on lattice gauge
theory with a non-positive weight in the partition function.   
\vspace{0.025cm}

{\bf Acknowledgments:} It is a pleasure to thank 
J.T. Lenaghan, P.H. Damgaard, F. Sannino, A.D. Jackson,
J.J.M. Verbaarschot, T. Sch\"afer, R. Pisarski, E.V. Shuryak,
M. Creutz, F. Wilczek, and M. Prakash for suggestions, comments, and 
criticism. This work was supported by U.S. DOE grant
DE-FG02-88ER40388.


\begin{thebibliography}{9}

\bibitem{QCDineq} D. Weingarten, Phys.~Rev.~Lett. {\bf 51} (1983) 1830;
  S. Nussinov, Phys. Rev. Lett. {\bf 51} (1983) 2081; E. Witten,
  Phys. Rev. Lett. {\bf 51} (1983) 2351. S. Nussinov and M. A. Lampert,
  Phys.Rept. {\bf 362} (2002) 193.

\bibitem{VW} 
C. Vafa and E. Witten, Nucl.~Phys. {\bf B 234} (1984) 173.

\bibitem{Elitzur}
S. Elitzur, Phys.~Rev. {\bf D 12} (1975) 3978.

\bibitem{AFG}
G.F. De Angelis, D. de Falco, and F. Guerra, Phys.~Rev. {\bf D 17} (1978)
1624.

\bibitem{AFGM}
G.F. De Angelis, D. de Falco, F. Guerra, and R. Marra, Acta Physica
Austriaca, Suppl. {\bf XIX} (1978) 205.


\bibitem{K}
P. Kosinski, Phys.~Rev.~Lett. {\bf 52} (1984) 1473.

\bibitem{FMS}
J. Fr\"ohlich, G. Morchio, and F. Strocchi, Nucl.~Phys. {\bf B 190}
(1981) 553.


\bibitem{footnote}
Notation: 
the index $i$ runs $i=1,\ldots,N$ and $\um=1,\ldots, d$ so
that $\prod_{i,\um}\equiv\prod_{i=1}^N\prod_{\um=1}^d$. Also, we use
the abbreviations d$\{A\}\equiv \prod_{i=1}^N\prod_{\um=1}^d A_{i,\um}$
and d$\{\phi\}\equiv \prod_{i=1}^N {\rm d}\phi_i$.


\bibitem{BDI} R. Balian, J. M. Drouffe, and C. Itzykson, Phys. Rev. {\bf D
    10} (1974) 3376.

\bibitem{kogut} J.B. Kogut, Rev. Mod. Phys, {\bf 51} (1979) 659. 


\bibitem{HJV} M.A. Halasz, A.D. Jackson, and J.J.M. Verbaarschot,
Phys. Lett. {\bf B 395} (1997) 293.


\bibitem{Langevin} J. Ambj{\o}rn, M. Flensburg, and C. Peterson, 
Phys. Lett. {\bf B 159} (1985) 335; J. Flower, S. W. Otto, and S. Callahan, 
Phys. Rev. {\bf D 34} (1986) 598; J. Ambj{\o}rn, M. Flensburg, and
C. Peterson, Nucl. Phys. {\bf B 275} (1986) 375. 


\bibitem{Luscher}
M.~L\"uscher, ``Absence of spontaneous breaking in Hammiltonian lattice Gauge
theories'', DESY-77-16-mc. 



\end{thebibliography}
\end{document}